\documentclass[journal]{IEEEtran}
\usepackage{graphicx} 
\usepackage{cite}
\usepackage{amsmath,amssymb,amsfonts}
\usepackage{algorithmic}
\usepackage{textcomp}
\usepackage{xcolor}
\usepackage{url}
\usepackage{caption}
\usepackage[absolute,overlay]{textpos}

\begin{document}

\title{Reproducing the Acoustic Velocity Vectors in a Spherical Listening Region}

\author{Frank Jiarui Wang, Thushara D. Abhayapala, Jihui Aimee Zhang, Prasanga N. Samarasinghe
\thanks{Frank Jiarui Wang was supported by the HDR Fee Remission Scholarship from the Australian National University.}
\thanks{F. J. Wang, T. D. Abhayapala and P. N. Samarasinghe are with the Australian National University, Ian Ross Building, North Road, Acton ACT 2601, Australia (e-mail: u5879960@anu.edu.au).}
\thanks{J. A. Zhang is with Institute of Sound and Vibration Research, University of Southampton, Highfield, Southampton SO17 1BJ, United Kingdom.}}

\maketitle

\begin{textblock*}{18cm}(1cm,26.5cm) 
  \fbox{\begin{minipage}[c]{\textwidth}
        \centering\small{ \copyright 2024 IEEE. Personal use of this material is permitted. Permission 
from IEEE must be obtained for all other uses, in any current or future 
media, including reprinting/republishing this material for advertising or 
promotional purposes, creating new collective works, for resale or 
redistribution to servers or lists, or reuse of any copyrighted 
component of this work in other works.}\end{minipage}}
\end{textblock*}

\begin{abstract}
Acoustic velocity vectors (AVVs) are related to the human's perception of sound at low frequencies and are widely used in Ambisonics. This paper proposes a spatial sound field reproduction algorithm called velocity matching, which reproduces the AVVs in the spherical listening region by matching the AVVs' spherical harmonic coefficients. Using the sound field translation formula, the spherical harmonic coefficients of the AVVs are derived from the spherical harmonic coefficients of the pressure, which can be measured by a higher-order microphone array. Unlike algorithms that only control the AVVs at discrete sweet spots, the proposed velocity matching algorithm manipulates the AVVs in the whole spherical listening region and allows the listener to move beyond the sweet spots. Simulations show the proposed velocity matching algorithm accurately reproduces the AVVs in the spherical listening region and requires fewer number of loudspeakers than pressure matching algorithm. 
\end{abstract}

\textbf{
}

\section{Introduction}
Spatial sound field reproduction aims to synthesize the desired sound field in the listening region. In most cases, reproduction is achieved by accurately reconstructing the pressure. Pressure based methods include matching the pressure at a number of sweet spots \cite{Nelson1993}, wave field synthesis (WFS) \cite{BerkhoutWFS, Boone1995, AhrensWFSRe, BerryWFS, WinterWFS} and higher order Ambisonics (HOA) \cite{Ward2001, Betlehem2005, Mark2005}. A large number of loudspeakers are usually involved to achieve sufficient accuracy in the reconstructed pressure, which alone does not guarantee satisfactory perception \cite{SporsReview}.

Acoustic velocity vectors (AVVs) are essential to WFS due to the Kirchhoff-Helmholtz integral \cite{AhrensWFSRe}. Inspired by WFS, in \cite{JPVM} and \cite{JPVMPlus}, reproduction was achieved by matching the pressure and the AVVs at discrete control points on the boundary of the listening regions. There was also attempt at solely matching the AVVs on the boundary \cite{Shin2016}. Measuring the AVVs at multiple control points involves a complicated setup. Moreover, \cite{JPVM} and \cite{JPVMPlus} required a large number of loudspeakers, which could be impractical for home theater or small exhibition space. 

Perceptually motivated sound field reproduction creates the desired perceptual effects by using psychoacoustics, and has the advantage of requiring fewer channels due to its tolerance to lower accuracy in the reproduced pressure \cite{Huseyin2017}. AVVs are related to the human's perception of sound below 700 Hz \cite{Gerzon1992, gerzon1992G} and are relevant to the interaural phase difference. Recently, AVVs were combined with mixed-source expansion to create immersive perception over an enlarged region \cite{ZuoAppSci}. AVVs were also used to create the desired perception at non-central listening points \cite{Wang2017}. 

AVVs have been applied to reproduction at sweet spots. Gerzon's velocity vector $r_{\text{V}}$ is widely used in Ambisonics \cite{Gerzon1992, gerzon1992G, AmbisonicsBook, Arteaga2013}. A time-domain method that jointly controls the AVVs and the pressures at multiple sweet spots was derived in \cite{Wen2021} and \cite{Wen2023}. To enable listener's movement beyond the sweet spots, the AVVs in the whole listening region should be characterized. 

A spherical region can be treated as an integral of multiple concentric spherical surfaces. In \cite{ZuoVel}, the spherical harmonic (SH) coefficients of the AVVs on each concentric spherical surface (abbrev. \textbf{SHV-surf} coefficients) were derived from the SH coefficients of the pressure in the spherical region. The SH coefficients of the pressure were measured by a spherical microphone array \cite{Meyer2002, Thushara2002}. The SHV-surf coefficients were of the form $X_{a}^{d}(k, r_{b})$ in which $r_{b}$ was the radius of the spherical surface on which the AVVs were characterized \cite{ZuoVel, ZuoAppSci}. SHV-surf coefficients were applied to virtual microphone synthesis to simulate the signals at microphones placed on an open sphere \cite{Erdem2023}. To compute the AVVs in a spherical region, the SHV-surf coefficients must be calculated for multiple radii. 

To characterize the AVVs in a spherical listening region using one set of SH coefficients across all radii, this paper presents the radial independent SH coefficients of the AVVs (abbrev. \textbf{SHV-indR} coefficients). The SHV-indR coefficients are of the form $({\zeta_{\boldsymbol{\hat{e}}}})_{a}^{d}(k)$ and the radial dependency is captured by a separate radial function $R_{a}(kr_{b})$. Starting from the definition of the AVVs, it is found that the SHV-indR coefficients are related to the SH coefficients of the pressure in the spherical listening region by the sound field translation formula. Therefore, the SHV-indR coefficients can be derived from spherical microphone array measurements. Unlike SHV-surf coefficients, which must be calculated at different radii, one set of SHV-indR coefficients is sufficient to characterize the AVVs throughout the spherical listening region. In \cite{Herzog2021}, the SHV-indR coefficients were derived by using eigenbeam-ESPRIT, which is a source localization method. 

This paper also proposes the velocity matching (VM) algorithm, which reproduces the desired AVVs in the spherical listening region by matching the SHV-indR coefficients. Due to the radial-independence, only one set of SHV-indR coefficients needs to be matched, whereas the SHV-surf coefficients must be matched on multiple concentric spherical surfaces. Simulation shows that VM accurately reproduces the AVVs at low frequencies and requires fewer number of loudspeakers than pressure based method.  

\section{AVV in the spherical region}
\begin{figure}[t]
  \centering
  \includegraphics[trim = 40mm 25mm 40mm 31mm, clip, width = 0.55\columnwidth]{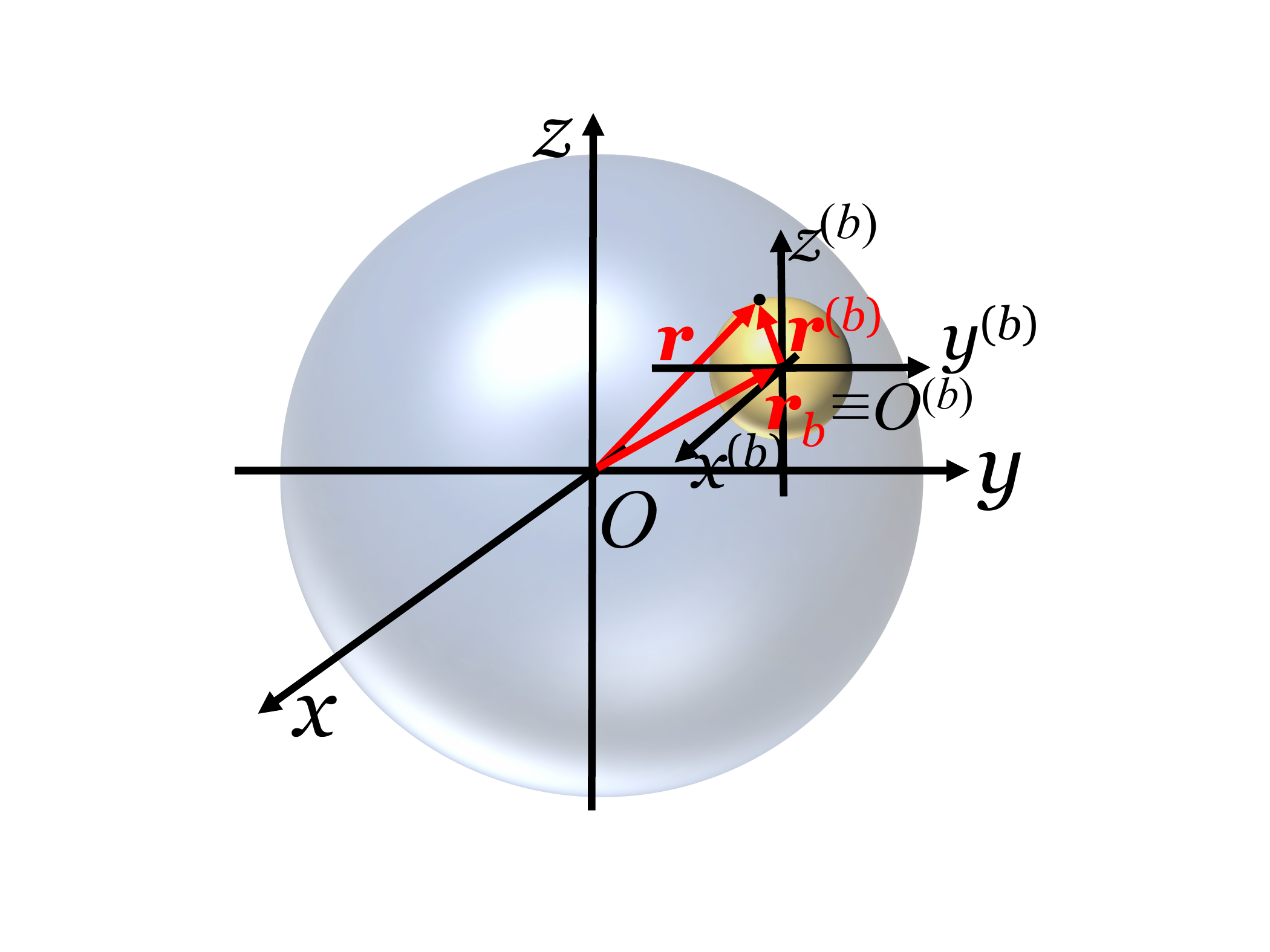}
  \vskip -4mm
  \caption{Setup of the geometric model. The listening region is in light blue. $\mathbf{r}_{b}$ is a point in the listening region.}
  \label{Fig:model_setup}
  \vskip -5mm
\end{figure}
\subsection{AVVs at a point}
\label{Sec:velocity_at_a_point}
Figure 1 shows the setup of the geometric model. The spherical listening region in light blue is free from sources and scatterers. The derivation starts from finding the AVVs at a point $\mathbf{r}_{b}$ within the listening region. The local $x^{(b)}y^{(b)}z^{(b)}$ coordinate system is centered at $\mathbf{r}_{b} \equiv O^{(b)}$. The $x^{(b)}y^{(b)}z^{(b)}$ coordinate system is the translation of the $xyz$ coordinate system with $\mathbf{r}_{b}$ as the translation vector. Note that $\mathbf{r} = \mathbf{r}_{b} + \mathbf{r}^{(b)}$. The superscript indicates the coordinate system used to express the location. If there are no superscripts, then the location is expressed with respect to the $xyz$ coordinate system. When using spherical coordinate system, $\mathbf{r} = (r, \theta, \phi)$ in which $r = ||\mathbf{r}||$, $\theta$ is the colatitude measured downward from the zenith (the positive $z$-axis), and $\phi$ is the azimuth measured counterclockwise from the positive $x$-axis in the $xy$ plane.

In Figure \ref{Fig:model_setup}, the pressure at a point $\mathbf{r}^{(b)} \equiv (r^{(b)}, \theta^{(b)}, \phi^{(b)})$ within the local region in yellow is
\begin{equation}
\label{Eq:local_SH}
p(k, \mathbf{r}^{(b)}) = \sum_{n=0}^{N} \sum_{m=-n}^{n} \beta_{n}^{m}(k, \mathbf{r}_{b}) j_{n}(kr^{(b)}) Y_{n}^{m} (\theta^{(b)}, \phi^{(b)})
\end{equation}
in which $k$ is the wavenumber, $j_{n}(\cdot)$ is the spherical Bessel function of the first kind, $Y_{n}^{m}(\cdot \,, \cdot)$ is the SH function of degree $n$ and order $m$, and $N$ is the truncation order. The SH coefficients $\beta_{n}^{m}(k, \mathbf{r}_{b})$ depend on the location of $\mathbf{r}_{b}$, which is the origin of the local coordinate system. From \cite{Fei2021}, the derivative $[\partial j_{n}(kr^{(b)})/\partial r^{(b)}]_{r^{(b)}=0} = (1/3) k \delta_{n, 1}$ in which $\delta_{n, 1}$ is the Kronecker delta function.  

Let $\rho_{0}$ denote the density of the medium and $c$ denote the speed of sound. Let $\boldsymbol{\hat{x}}$, $\boldsymbol{\hat{y}}$ and $\boldsymbol{\hat{z}}$ denote the unit vectors along the $x, y$ and $z$ axes, respectively. The AVV at $\mathbf{r}_{b} \equiv O^{(b)}$ is the linear combination of $\beta_{1}^{m}(k, \mathbf{r}_{b})$ \cite{Moore2017} 
\begin{align}
\label{Eq:local_vel_x}
    &V_{\boldsymbol{\hat{x}}}(\mathbf{r}_{b}, k) = \frac{i}{k\rho_{0}c}\frac{\partial p(k, \mathbf{r}^{(b)})}{\partial x} \bigg\vert_{r^{(b)}=0}\nonumber\\
    & =\sum_{n=0}^{N} \sum_{m=-n}^{n} \beta_{n}^{m}(k, \mathbf{r}_{b}) \frac{\partial j_{n}(kr^{(b)})}{\partial r^{(b)}} \bigg\vert_{r^{(b)}=0} Y_{n}^{m} \bigg(\frac{\pi}{2}, 0\bigg)\nonumber\\
    &= \frac{1}{3} \frac{i}{\rho_{0}c} \bigg[\sqrt{\frac{3}{8\pi}} \beta_{1}^{-1}(k, \mathbf{r}_{b}) - \sqrt{\frac{3}{8\pi}} \beta_{1}^{1}(k, \mathbf{r}_{b})\bigg],
\end{align}
\begin{align}
\label{Eq:local_vel_y}
    &V_{\boldsymbol{\hat{y}}}(\mathbf{r}_{b}, k) = \frac{i}{k\rho_{0}c}\frac{\partial p(k, \mathbf{r}^{(b)})}{\partial y} \bigg\vert_{r^{(b)}=0}\nonumber\\
    & =\sum_{n=0}^{N} \sum_{m=-n}^{n} \beta_{n}^{m}(k, \mathbf{r}_{b}) \frac{\partial j_{n}(kr^{(b)})}{\partial r^{(b)}} \bigg\vert_{r^{(b)}=0} Y_{n}^{m} \bigg(\frac{\pi}{2}, \frac{\pi}{2}\bigg)\nonumber\\
    &= \frac{1}{3} \frac{1}{\rho_{0}c} \bigg[\sqrt{\frac{3}{8\pi}} \beta_{1}^{-1}(k, \mathbf{r}_{b}) + \sqrt{\frac{3}{8\pi}} \beta_{1}^{1}(k, \mathbf{r}_{b})\bigg],\\
\label{Eq:local_vel_z}
    &V_{\boldsymbol{\hat{z}}}(\mathbf{r}_{b}, k) = \frac{i}{k\rho_{0}c}\frac{\partial p(k, \mathbf{r}^{(b)})}{\partial z} \bigg\vert_{r^{(b)}=0}\nonumber\\
    & =\sum_{n=0}^{N} \sum_{m=-n}^{n} \beta_{n}^{m}(k, \mathbf{r}_{b}) \frac{\partial j_{n}(kr^{(b)})}{\partial r^{(b)}} \bigg\vert_{r^{(b)}=0} Y_{n}^{m} (0, 0)\nonumber\\
    &= \frac{1}{3} \frac{i}{\rho_{0}c} \sqrt{\frac{3}{4\pi}} \beta_{1}^{0}(k, \mathbf{r}_{b}).
\end{align}
\subsection{The SHV-indR coefficients}
In Figure \ref{Fig:model_setup}, using the global $xyz$ coordinate system, the pressure at $\mathbf{r} \equiv (r, \theta, \phi)$ is 
\begin{equation}
    p(k, \mathbf{r}) = \sum_{\ell=0}^{L} \sum_{q = -\ell}^{\ell} \xi_{\ell}^{q}(k) j_{\ell}(kr) Y_{\ell}^{q} (\theta, \phi).
\end{equation}
in which $\xi_{\ell}^{q}(k)$ denotes the SH coefficients of the  pressure in the spherical listening region, and $L$ is the truncation order. Using the sound field translation formula \cite{SMABook}, 
\begin{align}
\label{Eq:translation_formula}
     &p(k, \mathbf{r}^{(b)}) = \sum_{n=0}^{N}  \sum_{m=-n}^{n} j_{n}(kr^{(b)}) Y_{n}^{m} (\theta^{(b)}, \phi^{(b)}) \nonumber\\ &\underbrace{\sum_{a = 0}^{A} \sum_{\ell=0}^{L} \sum_{q = -\ell}^{\ell} \xi_{\ell}^{q}(k) G_{nm}^{\ell q a}j_{a}(kr_{b}) Y_{a}^{(q-m)} (\theta_{b}, \phi_{b})}_{\beta_{n}^{m}(k, \mathbf{r}_{b}) \text{ in \eqref{Eq:local_SH}}} 
\end{align}
As shown in \eqref{Eq:local_vel_x}, \eqref{Eq:local_vel_y} and \eqref{Eq:local_vel_z}, since the AVVs involve only $\beta_{1}^{m}(k, \mathbf{r}_{b})$, the derivation restricts $n=1$ and $m = \{-1, 0, 1\}$. The term 
\begin{equation}
    G_{1m}^{\ell q a} = 4\pi i^{1+a-\ell} (-1)^{q} \sqrt{\frac{3(2\ell+1)(2a+1)}{4\pi}} W_{1} W_{2}
\end{equation}
in which $W_{1}$ and $W_{2}$ are the Wigner-3$j$ symbols \cite{KennedyBook}
\begin{equation}
    W_{1}=\begin{pmatrix}
        \ell & 1 & a\\
        0 & 0 & 0
        \end{pmatrix} \;\;\;
    W_{2}= \begin{pmatrix}
        \ell & 1 & a\\
        -q & m & q-m
        \end{pmatrix} .
\end{equation}
Let $d = q-m$, $\beta_{1}^{m}(k, \mathbf{r}_{b})$ in \eqref{Eq:translation_formula} becomes
\begin{align}
\label{Eq:global_SH_to_local_SH}
    &\beta_{1}^{m}(k, \mathbf{r}_{b}) \nonumber\\ &= 
    \sum_{a = 0}^{A} \sum_{\ell=0}^{L} \sum_{d = -\ell-m}^{\ell-m} \xi_{\ell}^{(d+m)}(k) G_{1m}^{\ell (d+m) a}j_{a}(kr_{b}) Y_{a}^{d} (\theta_{b}, \phi_{b}) \nonumber\\
    &=\sum_{a = 0}^{A} \sum_{d = -L-m}^{L-m} \underbrace{\bigg[\sum_{\ell =|d+m|}^{L} \xi_{\ell}^{(d+m)}(k) G_{1m}^{\ell (d+m) a}\bigg]}_{(\gamma_{1}^{m})_{a}^{d}} \nonumber\\
   & \qquad j_{a}(kr_{b}) Y_{a}^{d} (\theta_{b}, \phi_{b}) 
\end{align}
in which $A$ is the truncation order. 

Operator matrices can be constructed to link the SH coefficients  $\xi_{\ell}^{(d+m)}(k)$ of the pressure to $(\gamma_{1}^{m})_{a}^{d}(k)$ with $m \in \{-1, 0, 1\}$ such that
\begin{equation}
    (\pmb{\gamma}_{1}^{m})(k) = \mathfrak{B}_{1}^{m} \boldsymbol{\xi}(k)
\end{equation}
in which $(\pmb{\gamma}_{1}^{m})(k)$ and $\boldsymbol{\xi}(k)$ are the column vectors formed by concatenating $(\gamma_{1}^{m})_{a}^{d}(k)$ and $\xi_{\ell}^{(d+m)}(k)$, respectively. The operator matrices do not depend on the wavenumber $k$ (also the frequency) because $G_{1m}^{\ell (d+m) a}$ are frequency independent. 

The calculation of $\mathfrak{B}_{1}^{m}$ does not require significant resources because only three operator matrices with $m = \{-1, 0, 1\}$ are required. Moreover, $G_{1m}^{\ell (d+m) a}$ is non-zero only when $ |\ell-1| \leq a \leq \ell + 1$ \cite{KennedyBook}. Furthermore, since $W_{1} = 0$ when $a = \ell$ \cite{WignerWeb}, only two conditions $a = |\ell-1|$ and $a = \ell+1$ need to be considered. The dimension of $\mathfrak{B}_{1}^{m}$ is $L^2$ by $(L+1)^2$. This is because if $\xi_{\ell}^{(d+m)}(k)$ is measured up to degree $L$, the maximum degree of $(\gamma_{1}^{m})_{a}^{d}(k)$ can be calculated is $(L-1)$. For $a = L$, $\xi_{\ell}^{(d+m)}(k)$ with $\ell = L+1$ should be measured. 

Substituting \eqref{Eq:global_SH_to_local_SH} into \eqref{Eq:local_vel_x}, \eqref{Eq:local_vel_y} and \eqref{Eq:local_vel_z},  
\begin{equation}
\label{Eq:SH_decomp_vel}
    V_{\boldsymbol{\hat{e}}} (\mathbf{r}_{b}, k) = 
\sum_{a = 0}^{A} \sum_{d = -a}^{a} ({\zeta_{\boldsymbol{\hat{e}}}})_{a}^{d}(k) j_{a}(kr_{b}) Y_{a}^{d} (\theta_{b}, \phi_{b})
\end{equation}
in which $\boldsymbol{\hat{e}}\in \{\boldsymbol{\hat{x}}, \boldsymbol{\hat{y}}, \boldsymbol{\hat{z}}\}$ and $({\zeta_{\boldsymbol{\hat{e}}}})_{a}^{d}(k)$ denotes the SHV-indR coefficients of the form
\begin{align}
\label{Eq:SH_vel_x}
    ({\zeta_{\boldsymbol{\hat{x}}}})_{a}^{d}(k) &= \frac{1}{3} \frac{i}{\rho_{0}c} \bigg[\sqrt{\frac{3}{8\pi}}(\gamma_{1}^{-1})_{a}^{d}(k) - \sqrt{\frac{3}{8\pi}} (\gamma_{1}^{1})_{a}^{d}(k)\bigg],\\
\label{Eq:SH_vel_y}
    ({\zeta_{\boldsymbol{\hat{y}}}})_{a}^{d}(k) &= \frac{1}{3} \frac{1}{\rho_{0}c} \bigg[\sqrt{\frac{3}{8\pi}} (\gamma_{1}^{-1})_{a}^{d}(k) + \sqrt{\frac{3}{8\pi}} (\gamma_{1}^{1})_{a}^{d}(k)\bigg],\\
\label{Eq:SH_vel_z}
    ({\zeta_{\boldsymbol{\hat{z}}}})_{a}^{d}(k) &= \frac{1}{3} \frac{i}{\rho_{0}c} \sqrt{\frac{3}{4\pi}} (\gamma_{1}^{0})_{a}^{d}(k).
\end{align}

Operator matrices $\mathfrak{B}_{\boldsymbol{\hat{e}}}$ with $\boldsymbol{\hat{e}}\in \{\boldsymbol{\hat{x}}, \boldsymbol{\hat{y}}, \boldsymbol{\hat{z}}\}$ that compute the SHV-indR coefficients  $(\zeta_{\boldsymbol{\hat{e}}})_{a}^{d}(k)$ from the SH coefficients $\xi_{\ell}^{(d+m)}(k)$ of the pressure are constructed so that
\begin{equation}
\label{Eq:operator_mtx_vel}
    \boldsymbol{\zeta}_{\boldsymbol{\hat{e}}}(k) = \mathfrak{B}_{\boldsymbol{\hat{e}}} \boldsymbol{\xi}(k)
\end{equation}
in which $\boldsymbol{\zeta}_{\boldsymbol{\hat{e}}}(k)$ is the column vector formed by concatenating $(\zeta_{\boldsymbol{\hat{e}}})_{a}^{d}(k)$. From \eqref{Eq:SH_vel_x}, \eqref{Eq:SH_vel_y} and \eqref{Eq:SH_vel_z},
\begin{align}
    \mathfrak{B}_{\boldsymbol{\hat{x}}} &= \frac{1}{3} \frac{i}{\rho_{0}c} \bigg[\sqrt{\frac{3}{8\pi}} \mathfrak{B}_{1}^{-1} - \sqrt{\frac{3}{8\pi}} \mathfrak{B}_{1}^{1}\bigg],\\
    \mathfrak{B}_{\boldsymbol{\hat{y}}} &= \frac{1}{3} \frac{1}{\rho_{0}c} \bigg[\sqrt{\frac{3}{8\pi}} \mathfrak{B}_{1}^{-1} + \sqrt{\frac{3}{8\pi}} \mathfrak{B}_{1}^{1}\bigg],\\
    \mathfrak{B}_{\boldsymbol{\hat{z}}} &= \frac{1}{3} \frac{i}{\rho_{0}c} \sqrt{\frac{3}{4\pi}} \mathfrak{B}_{1}^{0}.
\end{align}

\subsection{Illustration of the AVVs in a spherical region}
\begin{figure}[t]
  \centering
   \includegraphics[trim = 17mm 40mm 20mm 50mm, clip, width = 0.8\columnwidth]{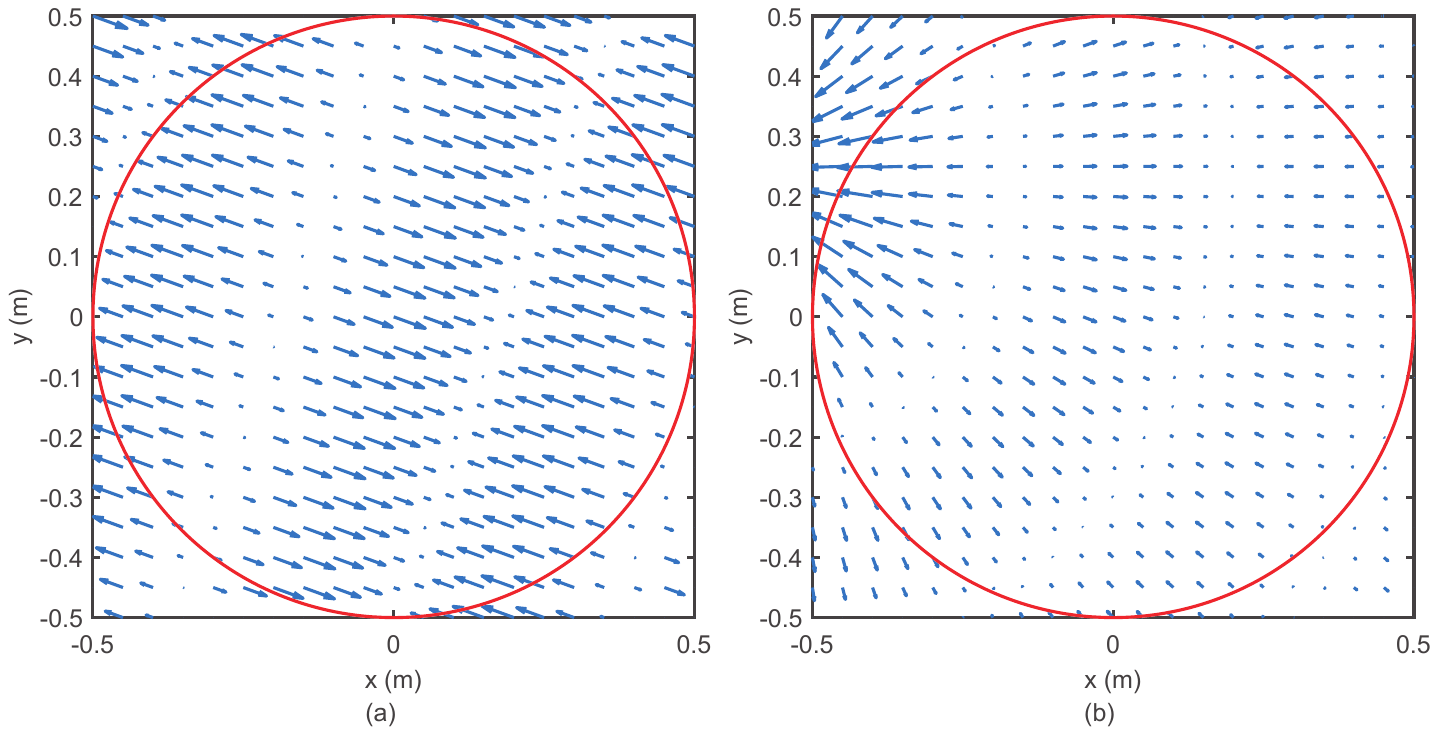}
   \vskip -3mm
   \caption{Real part of the AVVs on the $xy$ plane at 500 Hz. (a) A plane wave with incident direction $(\theta_{\text{pw}}, \phi_{\text{pw}}) =(\pi/2, 8\pi/9)$ rad. (b) A point source at $\mathbf{r}_{\text{ps}} = (0.7 \text{ m}, \pi/2 \text{ rad}, 8\pi/9 \text{ rad})$. The red circle with radius 0.5 m is the cross section of the boundary of the spherical listening region.}
   \label{Fig:vel_illu}
   \vskip -5mm
\end{figure}
For a plane wave with incident direction $(\theta_{\text{pw}}, \phi_{\text{pw}})$, 
\begin{equation}
    \xi_{\ell}^{q}(k) = 4\pi i^{\ell} \, \overline{Y_{\ell}^{q} (\theta_{\text{pw}}, \phi_{\text{pw}})} \quad \forall k.
\end{equation}
in which $\overline{(\cdot)}$ denotes conjugation. For a point source located at $\mathbf{r}_{\text{ps}} \equiv (r_{\text{ps}}, \theta_{\text{ps}}, \phi_{\text{ps}})$, 
\begin{equation}
    \xi_{\ell}^{q}(k) = -ik h_{\ell}^{(2)}(kr_{\text{ps}}) \, \overline{Y_{\ell}^{q} (\theta_{\text{\text{ps}}}, \phi_{\text{\text{ps}}})}
\end{equation}
in which $h_{\ell}^{(2)}(\cdot)$ is the spherical Hankel function of the second kind. Figure \ref{Fig:vel_illu} shows the real part of the AVVs on the $xy$ plane when the source is (a) a plane wave with incident direction $(\theta_{\text{pw}}, \phi_{\text{pw}}) =(\pi/2, 8\pi/9)$ rad, and (b) a point source at $\mathbf{r}_{\text{ps}} = (0.7 \text{ m}, \pi/2 \text{ rad}, 8\pi/9 \text{ rad})$. The red circle with radius 0.5 m is the cross section of the boundary of the spherical listening region. The SH coefficients $\xi_{\ell}^{q}(k)$ of the pressure are truncated to $L = 7$ and the SHV-indR coefficients $(\zeta_{\boldsymbol{\hat{e}}})_{a}^{d}(k)$ are truncated to $A = 6$. The density of the medium $\rho_{0} = 1.2042\; \text{kg/m}^{3}$ and the speed of sound $c = 343.21$ m/s. In Figure \ref{Fig:vel_illu}(a), the AVVs are either parallel or anti-parallel. In Figure \ref{Fig:vel_illu}(b), the AVVs either converge to or diverge from a point in the direction of $\phi = 8\pi/9$ rad and the spherical wave fronts of a point source can be discerned. 

\section{Reproducing the AVVs in a spherical region}
This section presents the velocity matching (VM) algorithm, which reproduces the desired AVVs in the spherical listening region by matching the SHV-indR coefficients. Assume there are $S$ loudspeakers with index $s = 1, 2, \cdots, S$. To characterize the loudspeakers, the pressure mode transfer functions ${(\xi^{\text{L}s})}_{\ell}^{q}(k)$ are measured, which represent the SH coefficients of the pressure in the spherical listening region when the input to the $s$-th loudspeaker is a unit sinusoid at frequency $kc/2\pi$ Hz. Next, by using the operator matrices in \eqref{Eq:operator_mtx_vel}, the $s$-th loudspeaker's velocity mode transfer functions $({\zeta_{\boldsymbol{\hat{e}}}^{\text{L}s}})_{a}^{d}(k)$ are obtained. To characterize the desired sound field, the SH coefficients $(\xi^{\text{des}})_{\ell}^{q}(k)$ of the desired pressure in the spherical listening region are measured by a spherical microphone array. Next, by using the operator matrices in \eqref{Eq:operator_mtx_vel}, the desired SHV-indR coefficients $({\zeta_{\boldsymbol{\hat{e}}}^{\text{des}}})_{a}^{d}(k)$ are found. A system of equations is established 
\begin{equation}
\label{Eq:vel_matching}
    \boldsymbol{\zeta}^{\text{des}}(k) = \mathbf{H}(k) \mathbf{w}(k).
\end{equation}
In \eqref{Eq:vel_matching}, $\boldsymbol{\zeta}^{\text{des}}(k) = [\boldsymbol{\zeta}_{\boldsymbol{\hat{x}}}^{\text{des}}(k)^{\textit{T}}, \boldsymbol{\zeta}_{\boldsymbol{\hat{y}}}^{\text{des}}(k)^{\textit{T}},\boldsymbol{\zeta}_{\boldsymbol{\hat{z}}}^{\text{des}}(k)^{\textit{T}}]^{\textit{T}}$ in which $\boldsymbol{\zeta}_{\boldsymbol{\hat{e}}}^{\text{des}}(k)$ with $\boldsymbol{\hat{e}}\in \{\boldsymbol{\hat{x}}, \boldsymbol{\hat{y}}, \boldsymbol{\hat{z}}\}$ is the column vector formed by concatenating $(\zeta_{\boldsymbol{\hat{e}}}^{\text{des}})_{a}^{d}(k)$ and $(\cdot)^{\textit{T}}$ denotes matrix transpose. $\mathbf{H}(k) = [\boldsymbol{\zeta}^{\text{L}1}(k), \boldsymbol{\zeta}^{\text{L}2}(k),\cdots, \boldsymbol{\zeta}^{\text{L}S}(k)]$ and its $s$-th column $\boldsymbol{\zeta}^{\text{L}s}(k) = [\boldsymbol{\zeta}_{\boldsymbol{\hat{x}}}^{\text{L}s}(k)^{\textit{T}}, \boldsymbol{\zeta}_{\boldsymbol{\hat{y}}}^{\text{L}s}(k)^{\textit{T}},\boldsymbol{\zeta}_{\boldsymbol{\hat{z}}}^{\text{L}s}(k)^{\textit{T}}]^{\textit{T}}$ in which $\boldsymbol{\zeta}_{\boldsymbol{\hat{e}}}^{\text{Ls}}(k)$ with $\boldsymbol{\hat{e}}\in \{\boldsymbol{\hat{x}}, \boldsymbol{\hat{y}}, \boldsymbol{\hat{z}}\}$ is the column vector formed by concatenating $(\zeta_{\boldsymbol{\hat{e}}}^{\text{L}s})_{a}^{d}(k)$. The column vector $\mathbf{w}(k)$ contains the loudspeaker weights. The VM algorithm is compared with the pressure matching (PM) algorithm \cite{Ward2001}, which finds the loudspeaker weights by matching the SH coefficients of the pressure in the spherical listening region. PM has a system of equations  
\begin{equation}
\label{Eq:p_matching}
    \boldsymbol{\xi}^{\text{des}}(k) = \mathbf{G}(k) \mathbf{w}(k)
\end{equation}
in which $\boldsymbol{\xi}^{\text{des}}(k)$ is the column vector formed by concatenating ${(\xi^{\text{des}})}_{\ell}^{q}(k)$. The matrix $\mathbf{G} (k) = [{\boldsymbol{\xi}^{\text{L1}}}(k), {\boldsymbol{\xi}^{\text{L2}}}(k), \cdots, {\boldsymbol{\xi}^{\text{LS}}}(k)]$ in which the $s$-th column  $\boldsymbol{\xi}^{\text{L}s}(k)$ is formed by concatenating $(\xi^{\text{L}s})_{\ell}^{q}(k)$. In both \eqref{Eq:vel_matching} and \eqref{Eq:p_matching}, the loudspeaker weights $\mathbf{w}(k)$ are found by Moore-Penrose pseudoinverse.

\begin{figure}[t]
\vskip -1mm
  \centering
  \includegraphics[trim = 30mm 65mm 25mm 70mm, clip, width = 0.9\columnwidth]{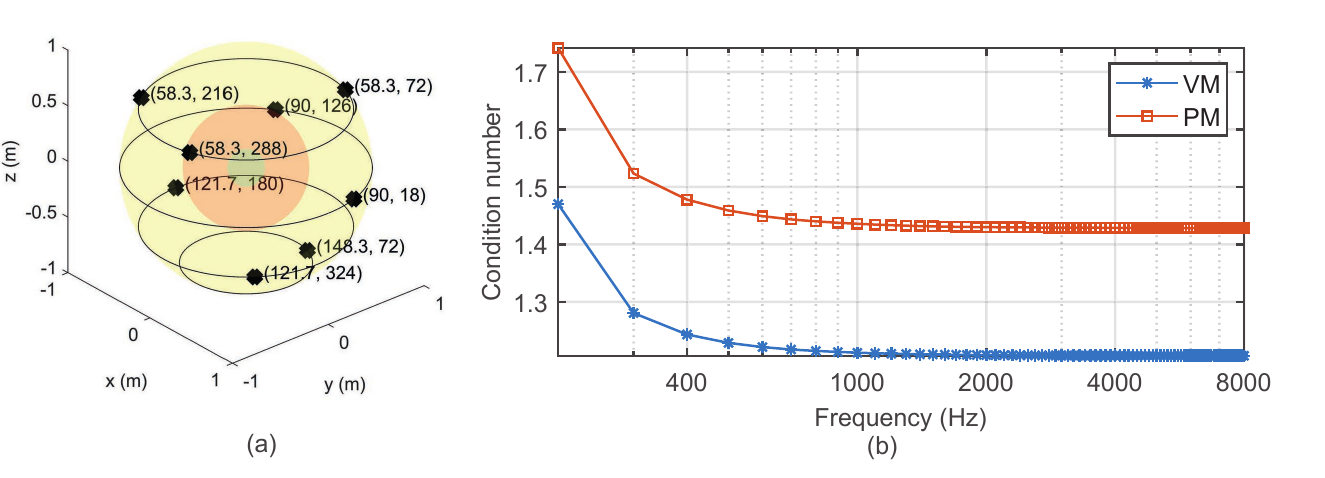}
  \vskip -3mm
  \caption{(a) Setup of the 8-channel loudspeaker array with loudspeakers denoted by black crosses. The loudspeakers are located on the yellow sphere with radius 1 m. The loudspeaker angular directions are denoted as $(\theta_{\text{L}s}, \phi_{\text{L}s})$ in degrees. The spherical listening region with radius 0.5 m is bounded by the red sphere. The cyan sphere at the center of the listening region has radius 0.15 m. (b) Condition numbers of $\mathbf{H}(k)$ for VM and $\mathbf{G}(k)$ for PM.}
  \vskip -3mm
  \label{Fig:reprod_setup}
  \vskip -2mm
\end{figure}

Figure \ref{Fig:reprod_setup}(a) shows the 8-channel loudspeaker array. The desired sound field is a plane wave with incident direction $(\theta_{\text{pw}}, \phi_{\text{pw}}) = (\pi/2, 8\pi/9)$ rad. The pressure SH coefficients $(\xi^{\text{des}})_{\ell}^{q}(k)$ and ${(\xi^{\text{L}s})}_{\ell}^{q}(k)$ are truncated to $\ell = 4$. Hence, the SHV-indR coefficients $({\zeta_{\boldsymbol{\hat{e}}}^{\text{des}}})_{a}^{d}(k)$ and $({\zeta_{\boldsymbol{\hat{e}}}^{\text{L}s}})_{a}^{d}(k)$ are truncated to $a = 3$. At each wavennumber $k$, the dimension of $\mathbf{H}(k)$ is 48-by-8 and the dimension of $\mathbf{G}(k)$ is 25-by-8. The density of the medium $\rho_{0} = 1.2042\; \text{kg/m}^{3}$ and the speed of sound $c = 343.21$ m/s. In Figure \ref{Fig:reprod_setup}(b), the condition numbers remain stable, though those of $\mathbf{H}(k)$ are slightly lower than those of $\mathbf{G}(k)$. The Moore-Penrose pseudoinverse is calculated by the \texttt{pinv} function in MATLAB and the default tolerance is used. Code for simulation can be accessed from \url{https://github.com/FJWang01/SH_Velocity}.

\begin{figure}[t]
  \centering
  \includegraphics[trim = 10mm 40mm 10mm 58mm, clip, width = 0.9\columnwidth]{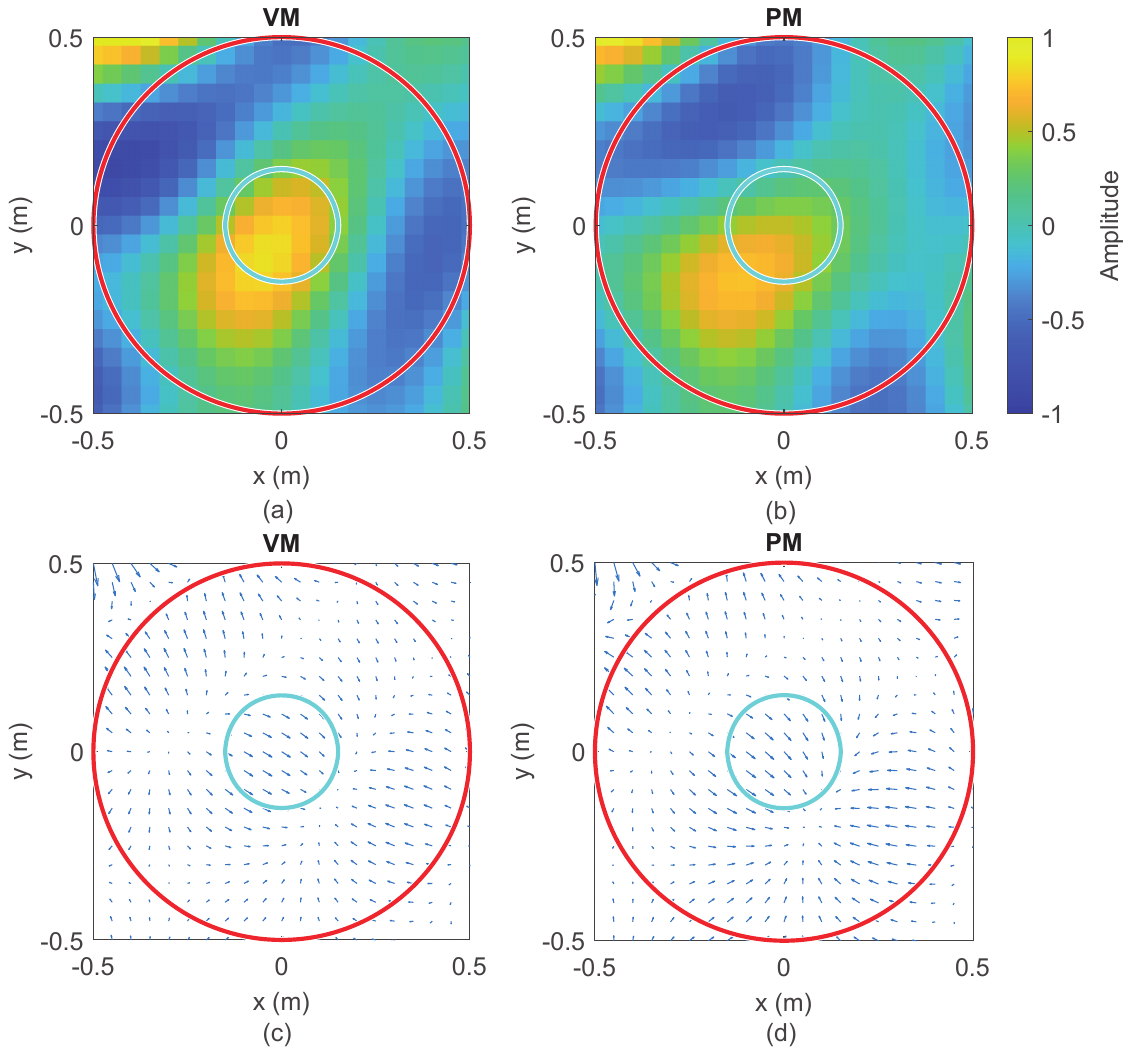}
  \vskip -5mm
  \caption{Real part of the reproduced pressure and the reproduced AVVs on the $xy$ plane at 500 Hz. The desired sound field is a plane wave with incident direction $(\pi/2, 8\pi/9)$ rad. (a) and (c) are reproduced by VM; (b) and (d) are reproduced by PM. The red circle and the cyan circle are the cross sections of the red sphere and the cyan sphere in Figure 3(a), respectively. }
  \vskip -1mm
  \label{Fig:reprod_160}
  \centering
  \includegraphics[trim = 40mm 70mm 40mm 73mm, clip, width = 0.9\columnwidth]{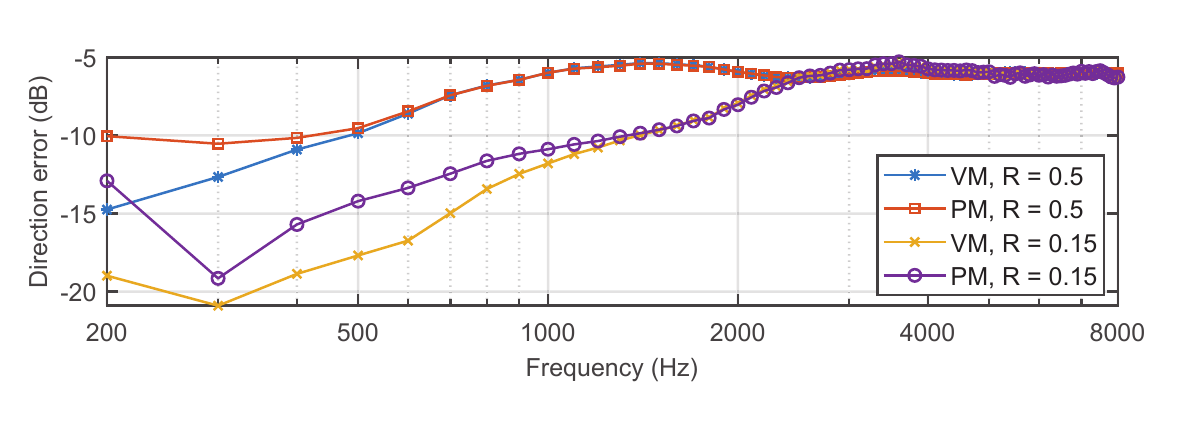}
  \vskip -6mm
  \caption{Velocity reproduction errors in the real part of the reproduced AVVs. R = 0.5: the spherical listening region bounded by the red sphere with radius 0.5 m in Figure 3(a); R = 0.15: the region bounded by the cyan sphere with radius 0.15 m in Figure 3(a).}
  \label{Fig:reprod_error}
  \vskip -5mm
\end{figure}

Figure \ref{Fig:reprod_160} shows the reproduced pressure and the AVVs on the $xy$ plane at 500 Hz. Figures \ref{Fig:reprod_160}(a) and \ref{Fig:reprod_160}(c) are reproduced by VM, whereas Figures \ref{Fig:reprod_160}(b) and \ref{Fig:reprod_160}(d) are reproduced by PM. The ground truth of the AVVs is in Figure \ref{Fig:vel_illu}(a). VM achieves better results in both reproduced pressure and reproduced AVVs. Like \cite{Lachlan2021} and \cite{ZuoIntensity}, the velocity reproduction error 
\begin{equation}
    \eta(k) = \cos^{-1} (\text{DOT}(k)) \; \text{rad},
\end{equation}
where
\begin{equation}
   \text{DOT}(k) = \frac{\mathbf{V}^{\text{des}} (\mathbf{r}_{b}, k)}{||\mathbf{V}^{\text{des}} (\mathbf{r}_{b}, k)||_{2}} \cdot \bigg[\frac{\mathbf{V}^{\text{re}} (\mathbf{r}_{b}, k)}{||\mathbf{V}^{\text{re}} (\mathbf{r}_{b}, k)||_{2}} \bigg]^{T}
\end{equation}
in which $\mathbf{V}^{\text{des}}(\mathbf{r}_{b}, k) \equiv [V_{\boldsymbol{\hat{x}}}^{\text{des}}(\mathbf{r}_{b}, k), V_{\boldsymbol{\hat{y}}}^{\text{des}}(\mathbf{r}_{b}, k), V_{\boldsymbol{\hat{z}}}^{\text{des}}(\mathbf{r}_{b}, k)]$ is the desired AVV and $\mathbf{V}^{\text{re}} (\mathbf{r}_{b}, k) \equiv [V_{\boldsymbol{\hat{x}}}^{\text{re}}(\mathbf{r}_{b}, k), V_{\boldsymbol{\hat{y}}}^{\text{re}}(\mathbf{r}_{b}, k), V_{\boldsymbol{\hat{z}}}^{\text{re}}(\mathbf{r}_{b}, k)]$ is the reproduced AVV. Here, only the real part of the AVVs are considered. In Figure \ref{Fig:reprod_error}, the blue line and the red line illustrate the velocity reproduction errors averaged across 113081 evaluation points within the red sphere with radius 0.5 m in Figure \ref{Fig:reprod_setup}(a). VM achieved lower errors below 500 Hz. Above 500 Hz, the errors of VM and PM are similar. The yellow line and the purple line show the velocity reproduction errors averaged across 2993 evaluation points within the cyan sphere of radius 0.15 m in Figure \ref{Fig:reprod_setup}(a). Up to slightly above 1 kHz, VM performs better than PM. The simulation result suggests that to reproduce the desired AVVs at low frequencies, VM requires fewer loudspeakers than PM. VM has strong potentials in home theater and small exhibition spaces, where space constraints only allow the installation of small loudspeaker array. For reproduction at mid to high frequencies, intensity based method \cite{ZuoIntensity, ZuoIntensityEU, ZuoIntensityMZ, Choi2004, Arteaga2013} can be used. 
 
\section{Conclusion}
This paper presented the SHV-indR coefficients, which were the radial independent SH coefficients of the AVVs in a spherical region. The SHV-indR coefficients were derived from the SH coefficients of the pressure in the spherical region by using the sound field translation formula. The SHV-indR coefficients were used in VM, which reproduced the AVVs in the spherical listening region by matching the SHV-indR coefficients. Simulation showed at low frequencies, VM accurately reproduced the AVVs using few number of loudspeakers. Future work will include conducting perceptual tests and investigating methods to enlarge the listening region.  

\bibliographystyle{IEEEtran}
\bibliography{refs23}

\end{document}